\documentclass[nofootinbib]{revtex4}
 \usepackage{graphicx}

\begin{document}

\title{\bf{A note on DSR}}

\author{\small \em Carlo Rovelli}
\affiliation{\small\it Centre de Physique Th\'eorique de Luminy\footnote{Unit\'e mixte de recherche (UMR 6207) du CNRS et des Universit\'es de Provence (Aix-Marseille I), de la M\'editerran\'ee (Aix-Marseille II) et du Sud (Toulon-Var); laboratoire affili\'e \`a la FRUMAM (FR 2291).}, Case 907, F-13288 Marseille, EU} 
\date{\small August 27, 2008}

\begin{abstract}
\noindent
I study the physical meaning of Deformed, or Doubly, Special Relativity (DSR). I argue that DSR could be physically relevant in a certain large-distance limit.  I consider a concrete physical effect: the gravitational slowing down of time due to the gravitational potential well of a massive-particle, and its effect on the dynamics of the particle itself. I argue that this physical effect can survive in a limit in which gravitation and quantum mechanics can be disregarded, and that taking it into account leads directy to the Girelli-Livine DSR formalism.  This provides a physical interpretation to the corresponding 5d spacetime, and a concrete physical derivation of DSR.
\end{abstract}
\maketitle

\section{The DSR physical regime}

\noindent Consider a physical system characterized by a length $l$, a time interval $t$ and mass $m$.   With these quantities, we can define a characteristic velocity $v=l/t$, a characteristic action $s=ml^2/t$, and other characteristic quantities such as $x=l^3/(mt^2)$, which has the same dimension as the Newton constant $G$.   In the 3d space of the physical regimes coordinatized by $t,l$ and $m$, the three fundamental constants $c$, $\hbar$ and $G$ --speed of light, Planck constant and Newton constant--  and their combinations, such as the Planck length $L_P=\sqrt{\hbar G/c^3}$ and Planck mass $M_p=c\sqrt{\hbar/G}$, set scales at which characteristic phenomena become relevant.  For instance, if $v\ll c$ we expect relativistic effects to be negligible and therefore nonrelativistic physics to apply; if $s\gg\hbar$ we expect quantum effects to be negligible and therefore classical  (non--quantum) physics to apply.  Similarly, if $x\gg G$ we expect that gravitational effects can be neglected.  In other words, classical general relativity (GR), Minkowski--space quantum field theory (QFT), classical special relativity, non--relativistic quantum mechanics and non--relativistic classical mechanics are approximate physical theories that describe well the world in certain asymptotic regions of the ``space of the physical regimes", coordinatized by  $l$, $t$ and $m$.  These asymptotic regions represent the physical regimes in which the various approximate theories apply.    Needless to say, we do not yet know the complete physical theory that describe all regimes in the $(l,t,m)$ space.

  \begin{figure}[h]
  \begin{center}
  \includegraphics[height=5cm]{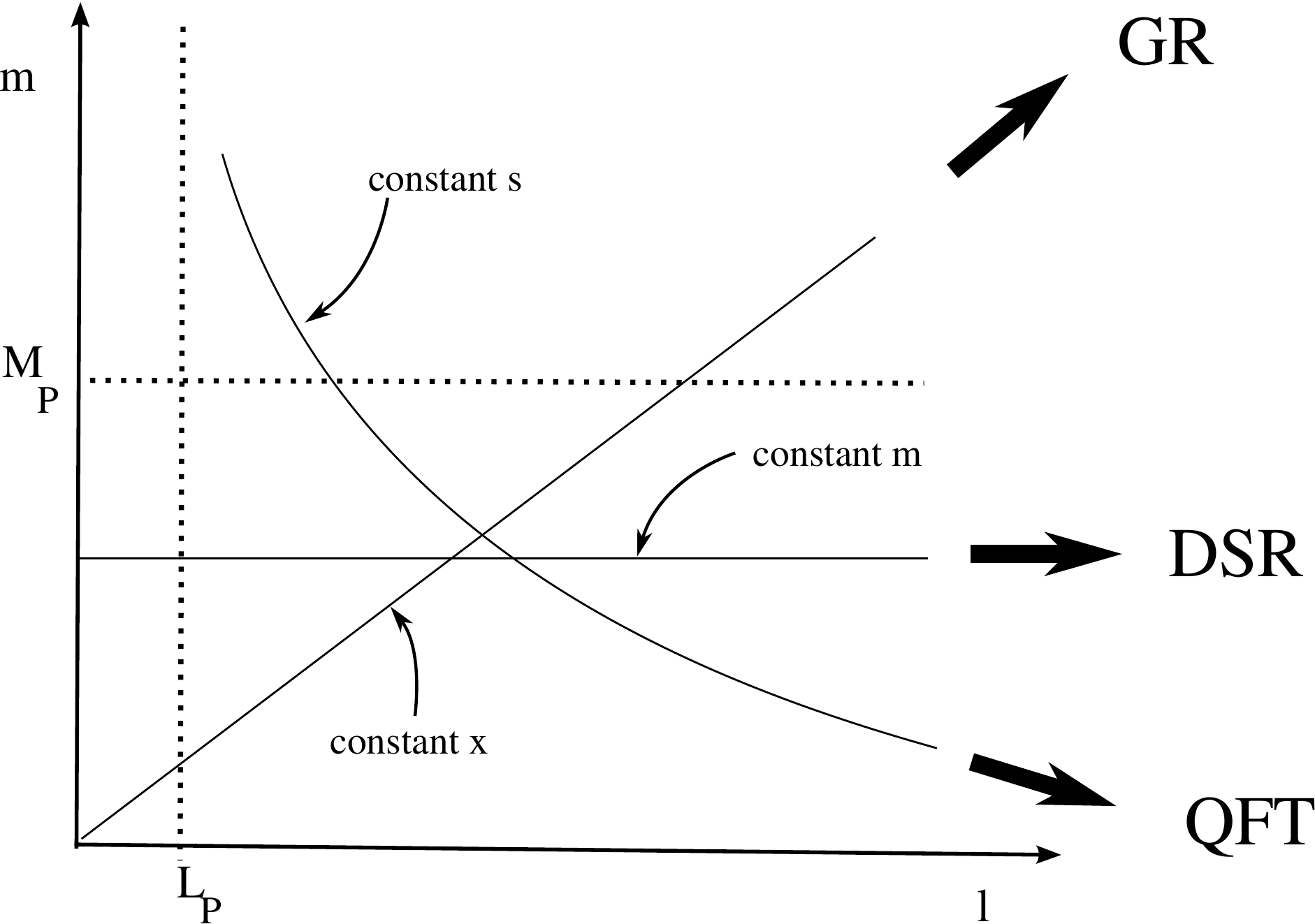}
  \end{center}
  \caption{Asymptotic regions in the space of the physical regimes, and theories describing them.}
 \end{figure}

Still, the asymptotic regions mentioned above are not necessarily the only ones that may have physical relevance for us.  Other asymptotic regions can be considered. To see this, let us
from now on consider only relativistic physics, namely regimes where $v$ is not necessarily small with respect to $c$,  choose units where $c=1$ and identify lengths and time units.  Then the space of the relativistic regimes is two dimensional, and can be parametrized by $l$ and $m$. Say we are interested in regimes where $l>>L_P$. Since $L\sim 10^{-33}cm$, this includes virtually all the regimes we have access to.  Now, the point I want to emphasize is that there are several distinct asymptotic large $l$ regions in the space of the physical regimes.

In particular, we can take the limit $l\gg L_P$ at constant $s=lm$, or at constant $x=l/m$.   In the first case, we get to a regime where $x$ is large (hence gravity can be neglected), and actions are arbitrary: this is the regime where QFT holds. (Notice that from this perspective QFT is a large distance regime: we do not test Planck length physics at CERN.)  In the second case,  we get to a regime where actions are large (hence quantum mechanics can be neglected), while gravity cannot be neglected.  This is the regime where GR holds.    But: {\em what happens if we take the large $l$ limit at constant $m$?} (See Figures 1 and 2.)

If we take the large $l$ limit at constant $m$, we get to a region where $s$ and $x$ are large, namely gravity and quantum theory can both be neglected, but $m$ is finite and can be of the order of the Planck mass.  In this regime we can have arbitrary finite masses $m$ at arbitrarily high densities $m/l^3$. Let us denote this regime as the ``relativistic high--density regime", or the DSR regime (from Deformed, or Doubly Special Relativity \cite{dsr}, or perhaps from high DenSity Relativistic regime).

\begin{figure}[h]
  \begin{center}
  \includegraphics[height=5cm]{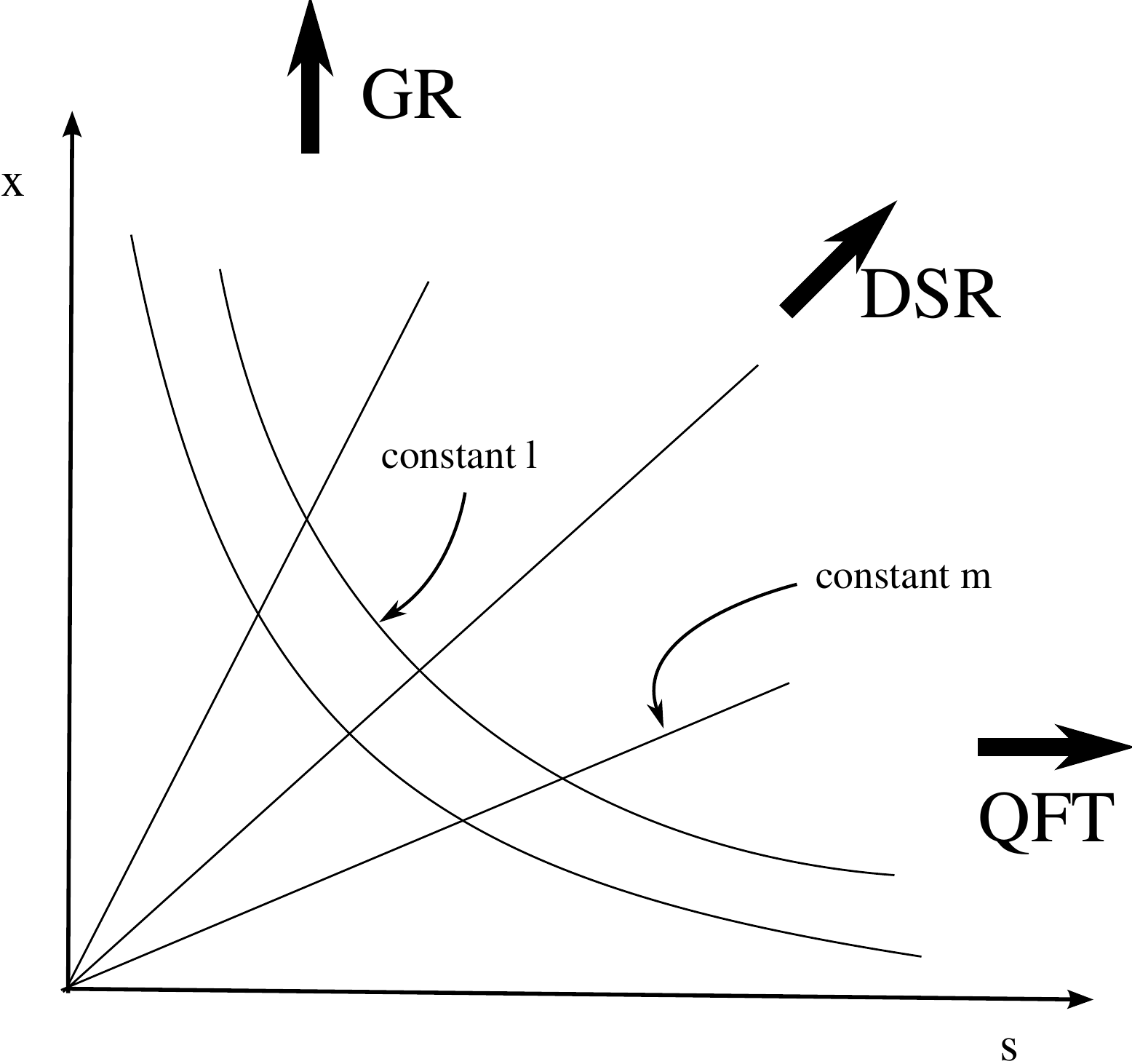}
  \end{center}
  \caption{\label{Boundary02} Same as figure 1, but in $x$, $s$ coordinates.}
  \end{figure}

The DSR asymptotic region is therefore a region that exists in the space of the physical regimes, where distances $l$ are large compared with the Planck length, actions $s$ are large compared to $\hbar$ and $x$ is large compared to $G$, but masses $m$ can be of the order of the Planck mass $M_P$, and densities can be arbitrarily large.  In other words it is a region where we can take the limit $\hbar\to0$ and $G\to0$, but only keeping constant the ratio $M_P^2=\hbar/G$.  Now the question is: is this simply a region described by classical special relativity?

The answer depends on whether or not there are real physical effects governed by the ratio $\hbar/G$.  If these exist, they play a role also in the DSR region, in spite of the fact that in a DSR region we can neglect purely quantum as well as purely gravitational phenomena.  

In the next section I investigate whether phenomena proportional to $\hbar/G$ exist in nature, and therefore whether we must expect a nontrivial DSR regime to exist.

\section{Mass slows clocks}

To get to the DSR regions, we need masses $m$ of the order of the Planck mass.  Since the Planck mass is of the order of micrograms,  at first sight classical special relativity appears to be perfectly capable of describing a system with large actions, small $x$ and masses of the order of a microgram.   However, this is true for \emph{extended} objects with the mass of a microgram.  In special relativity, on the other hand, we also consider  \emph{point} particles.  This is legitimate since we are interested in distance scales larger than $L_P$, hence we can forget any eventual internal structure of a particle, as far as we do not probe the Planck-length scale.  But are we sure that a   \emph{point}  particle with the mass close to the Planck mass (or an energy close to the Planck energy) is not affected by physical phenomena of the order of $\hbar/G$?

To answer this question with certainty, we would need to know full quantum gravity, compute in an arbitrary regime, and study the limit. We are not able to do so, but we can nevertheless assume that basic gravitational and quantum phenomena remain true at every scale, and work inductively on the basis of these.  Let us therefore do so, for a small particle of size $l$ and mass $m$. Let us call $s$ the proper time of the particle. Recall that the 4-momentum of the particle is given by $p^\mu=mdx^\mu/ds$. The proper time  $s$ is the time that an hypothetical clock on the particle itself would measure.  That is, it is the time interval ``felt" by the particle.

According to GR, proper time slows down if there is an energy density.  In the rest frame of the particle, the energy density is $m/l^3$. In the static approximation, this gives a metric
\begin{equation}
ds^2=(1+2\Phi(\vec x))dt^2-d\vec x^2
\end{equation}
where $\Phi(\vec x)$ is the Newtonian potential of the particle, which is $\Phi(\vec x)\sim - Gm/l$. Hence, at the particle and in its rest frame the metric is
\begin{equation}
ds^2=\left(1-\frac{2Gm}{l}\right)dt^2
\end{equation}
Quantum theory associates a length scale to any object of mass $m$: its Compton wavelength $\lambda_m=\hbar/m$. The Compton wavelength represents a natural quantum ``spread" of a point particle of mass $m$.  In the large distance limit in which we disregard the size of a point particle, we can reasonably conjecture that any short scale size effect would be in fact determined by its Compton wavelength.  In particular, we can  reasonably conjecture that its Newtonian potential can be taken to be $\Phi(\vec x)\sim - Gm/\lambda_m$, as if its mass was spread on a region of Compton-length size. It follows that the proper time $ds_m$ of the particle is related to the Minkowski proper time along the particle trajectory by
\begin{equation}
ds^2=\left(1-2\frac{Gm}{\lambda_m}\right)dt^2
=\left(1-2\frac{Gm}{\lambda_m}\right)dt^2
=\left(1-2\frac{Gm^2}{\hbar}\right)dt^2.
\end{equation}
Let us call $ds_0$ the special relativistic proper time along the worldline of the particle, computed disregarding GR. Then
\begin{equation}
ds_m= {\Gamma}^{-1}ds_0
\label{dsdt}
\end{equation}
where
\begin{equation}
\Gamma\equiv \sqrt{\frac{1}{1-2\frac{m^2}{M_P^2}}}.
\end{equation}
It is reasonable to assume that the 4-momentum that governs the dynamics of a high--mass relativistic point particle is not $p^\mu=mdx^\mu/ds_0$, but rather $p^\mu=mdx^\mu/ds$, because according to general relativity the particle itself, as far all physical phenomena are concerned, exists in a spacetime where proper temporal intervals are given by $ds$, not $ds_0$.

In other words, general relativity teaches us that energy creates negative gravitational potential, and this slows clocks down. This effect must have a consequence on  high mass relativistic point particles.

Notice that this physical effect depends on $\hbar/G$: therefore {\em it survives in the DSR regime} even when purely quantum effects and purely gravitational effects can be neglected.  We have therefore find a phenomenon with the features we were looking for.  A point particle whose energy is comparable with the Planck energy is likely to be affected by this phenomenon, and therefore do not obey simple special relativistic physics. Which theoretical framework is likely to describe it?

\section{DSR}

The immediate effect of (\ref{dsdt}) is that the 4-momentum of the particle becomes
\begin{equation}
p^\mu=m\frac{dx^\mu}{ds}=\Gamma\ \frac{dx^\mu}{ds_0}= \Gamma \ p^\mu_0
\end{equation}
where $p^\mu_0$ is the conventional special relativistic momentum of the particle.
Let us define  the five-momentum
\begin{equation}
\pi^I=(\Gamma p^\mu_0, \Gamma\, M_P)= \left(\Gamma\,m\, \frac{dx^\mu}{ds_0},\Gamma\, M_P\right)
\end{equation}
where $I=0,1,2,3,4$, and observe that
\begin{equation}
\pi^I \pi^J\eta_{IJ}= M_P^2
\label{ppM}
\end{equation}
where
\begin{equation}
\eta_{IJ}= {\rm diag}[-1,1,1,1,1].
\end{equation}

A conventional special relativistic reference frame is uniquely determined by a single free massive point particle (plus something fixing the orientation of 3d space).   Given a particle $O$, let  $p_i^\mu$ be the  4-momentum of an ensemble of particles, labelled by an index $i$. If I use, instead, the particle $p_i$ to fix the reference system, I expect a Lorentz matrix $\Lambda$ to transform all 4-momenta $p_i^\mu$ into the 4-momenta $\tilde p_i^\mu=\Lambda^\mu{}_\nu  p_i^\nu$ relative to a frame determined by the second particle.

If we do not restrict ourselves to particles that have mass small compared to $M_p$, however, we must also take into account the mass of the particle, an the differences between the proper times of the particles.  If the particles have different masses, the relation between momenta should be affected by the mass.  The transformation must preserve the invariant quantity $M_P$. It is natural for it to be formed by $SO(4,1)$ transformations on $\pi^I$, which preserve the quadratic form  (\ref{ppM}).  Clearly, we obtain usual special relativity in the small $M_P$ limit, where we can take $\Gamma=1$.

This observation leads immediately to the Girelli-Livine DSR formalism \cite{FloEt}, which, in turn, is equivalent to several previous formulations of DSR physics \cite{dsr}.

\section{5d ``extended spacetime"}

The above considerations lead to an interpretation of the 5d spacetime that carries a natural formulation of the DSR mathematics. Fix a Lorentz frame and a relativistic massive particle with mass $m$. Let $x^0=t$ and $\vec x$ be standard Lorentz coordinates. Let $x^4=s_0$ be the proper time measured by a free particle of mass $m\ll M_P$. Let $\cal D$ be the five dimensional space spanned by these coordinates.   In special relativity, each point $x^\mu$ is reached by a free particle coming from the origin in a proper time $s_0^2=t^2-\vec x^2$.  This equations defines a light cone in $\cal D$ and the linear group that leaves this light cone invariant  is $SO(4,1)$.

\section{A bold speculation on $SO(4,1)$ transformations}

Consider now a particle of mass $\tilde m$ moving on the same Minkowski space. Its proper time is given by $\tilde s=\Gamma^{-1}s_0$.
It is tempting to conjecture that if we use rods and clocks all made by particles of mass $\tilde m$, then, \emph{in general}, we \emph{still} measure distances $\vec{\tilde x}$ and times $\tilde t$ forming a special relativistic spacetime satisfying
 $\tilde s^2=\tilde t^2-\vec{\tilde x}^2$, where, however,
\begin{eqnarray}
\tilde t&=& \Gamma(t-\Gamma s) \\
\tilde s&=& \Gamma(s-\Gamma t).
\end{eqnarray}
Is this credible? If it was so, a particle of mass $m\ll M_P$ would appear as a particle with a mass comparable with $M_P$, seen by an ensemble of particles of masses of the order of $M_P$.  That is, mass would be a relative concept.

\section{Conclusions}

-- The physics conjectured in various DSR approaches can be understood as a simple consequences of GR and quantum theory. In a large distance regime, we can approximate particles as pointlike. Around a small particle, time slows down because of the gravitational potential. If we assume that quantum effects are equivalent to the spread of the gravitational source over a Compton wavelength, we obtain a local proper time associated to a massive particle different from Minkowski proper time by a factor that depends on $\hbar/G$, which survives in the limit in which both $G$ and $\hbar$ are taken to zero.

- The idea investigated here is that a localized particle produces a backreaction on
the metric, which in turns affects the dynamics of the propagating particle.  This idea
has been already explored in the literature, in the form of an energy-dependent
metric.   Here I have studied a concrete mechanism via which this energy dependence
can be realized:  the local slowing down of time due to the potential well of the particle.

-- Obviously the slowing down of the clocks considered here does not affect large extended objects.  We need the size to be (ideally) smaller than the Compton wavelength. Hence, it is the momentum of the idealized relativistic \emph{point} particles which is affected by this physical effect, not the momentum of extended objects.  Notice, indeed, that the observation made here  can be read as supporting the idea that what is relevant for DSR is the 
energy \emph{density} and not the total energy, a conclusion convincingly defended 
in particular by Sabine Hossenfelder \cite{Sabine}.

-- The DSR regime is a \emph{large distance} regime, in which we look at physics large compared to the Planck length.

-- The Girelli-Livine DSR formalism \cite{FloEt}, and therefore, in turn, several previous formulations of DSR physics
\cite{dsr}, follow naturally from this phenomenon.

-- Can we interpret the $SO(4,1)$ transformations as invariances of Nature?   Is there anything relative in the concept of mass?

\acknowledgements

\noindent Thanks to Florian Girelli and Etera Livine for explanations on DSR. The idea in this note emerged in discussions with Lee Smolin sometime ago.  I thank him for encouraging me to publish this old note.  The development of Lee's ideas on this subject, along lines related to what is done here, is in \cite{Lee}.


\begin{thebibliography}{99}


\bibitem{dsr} G Amelino-Camelia, ``Giovanni Amelino-Camelia, Relativity in space-times with short-distance structure governed by an observer-independent (Planckian) length scale", Int. J. Mod. Phys. D11 (2002) 35-60; ``Testable scenario for relativity with minimum-length", Phys. Lett. B 510 (2001) 255.  J Kowalski-Glikman, ``Introduction to Doubly Special Relativity", Lect. Notes Phys.669:131-159, 2005.  Joao Magueijo, Lee Smolin, ``Lorentz invariance with an invariant energy scale", Phys.Rev.Lett. 88 (2002) 190403; ``Generalized Lorentz invariance with an invariant energy scale", Phys.Rev. D67 (2003) 044017.

\bibitem{FloEt} F Girelli, ER Livine, ``Physics of Deformed Special Relativity: Relativity Principle revisited", gr-qc/0412004 and gr-qc/0412079.

\bibitem{Sabine} S Hossenfelder ``Multi-Particle States in Deformed Special Relativity", Phys. Rev. D 75, 105005 (2007). 

\bibitem{Lee} L Smolin, ``Could deformed special relativity naturally arise from the semiclassical limit of quantum gravity?" arXiv:0808.3765. 

\end{thebibliography}
\end{document}